\documentstyle[aasms4]{article}
\begin{document}

%-------------------------------------------------

\title{Distributed Burning in Type Ia Supernovae: A Statistical Approach}
\author{A.M. Lisewski, W. Hillebrandt}
\affil{Max\--Planck\--Institut f\"ur Astrophysik,
Karl-Schwarzschild\--Str. 1, 85740 Garching, Germany}
\author{S.E. Woosley}
\affil{UCO/Lick Observatory, University of California Santa Cruz,
Santa Cruz, CA 95064, USA}
\author{J.C. Niemeyer}
\affil{University of Chicago, Department of Astronomy and Astrophysics,
5640 S. Ellis Avenue, Chicago, IL 60637, USA}
\and
\author{A.R. Kerstein}
\affil{Combustion Research Facility, Sandia National Laboratories,
Livermore, CA 94551-0969, USA}

%-------------------------------------------------

\begin{abstract}

We present a statistical model which shows the influence of turbulence on a
thermonuclear flame propagating in C+O white dwarf matter. Based on a 
Monte Carlo description of turbulence, it provides a method for 
investigating the physics in the so-called distributed burning regime. 
Using this method we perform numerical simulations of turbulent
flames and  show that in this particular regime the flamelet model for
the turbulent 
flame velocity loses its validity. In fact, at high turbulent
intensities burning in the distributed
regime can lead to a deceleration
of the turbulent flame and thus induces a competing process to
turbulent effects that cause a higher flame speed. It is also 
shown that in dense C+O matter turbulent heat transport is described
adequately by the Peclet number, Pe, rather than by the
Reynolds number, which means that flame propagation is decoupled
from small-scale turbulence. Finally, at the onset of our results 
we argue that the available turbulent energy in an exploding C+O 
white dwarf is probably too low in order to make a deflagration to 
detonation  transition possible.
\end{abstract}


%-------------------------------------------------

\section{Introduction}

The thermonuclear explosion of a C+O Chandrasekhar-mass white dwarf,
which is believed to be the underlying process of a type Ia supernova
(SN Ia), has been subject to numerous investigations. However, despite
the fact that there are different plausible models explaining the
history of the explosion (\cite{nomoet84}; \cite{woowea86};
\cite{muellar86}; \cite{liv93}; \cite{khok91a}; \cite{arneliv94};
\cite{niemhill95}; \cite{hoefkhokwhee95}; \cite{hoef95};
\cite{wheeetal95}; \cite{hoefkhok96}), many important details still
remain unclear. Thermonuclear reactions provide the source of energy
which possibly unbinds the white dwarf. Thus, a SN Ia is characterized
by the physics of 
thermonuclear flames that propagate through the star. 
The physical conditions of this flame vary drastically during
the different temporal and spatial stages of this process. A better
understanding of these conditions, their interaction with the flame
and finally their consequences regarding the explosion itself are
still important issues from the theoretical point of view. 

Here, we focus on the interaction between thermonuclear burning and
turbulence taking place during the burning process. Turbulence is
caused by different kinds of instabilities (like shear instabilities
or the Rayleigh\--Taylor instability) that occur on certain length and
time scales (for an overview, see \cite{niemwoo97}). Rough estimates
give a turbulent Reynolds number Re $ \approx 10^{14}$ at an integral
scale of $L \approx 10^{6}\,$cm (\cite{hillniem97}). Consequently, the Kolmogorov
scale $ l_{\rm k}$, i.e. the scale where microscopic dissipation
becomes important, is about $ 10^{-4}\,$cm.  This enormous dynamical
range makes a direct representation -- at least by means of numerical
methods -- practically impossible. On the other hand, turbulence is a
characteristic feature during the explosion process and it must be
considered in any realistic model of a SN Ia. The simultaneous
coupling to energy generation due to nuclear reactions leads directly
to the physics of turbulent combustion, where a few crucial problems,
even for terrestrial conditions, still remain unsolved. These problems
include the prescription of the effective turbulent flame speed or  the
existence of different modes in turbulent combustion along with their
physical properties. Global features of turbulent combustion can be
systematically classified by a small set of dimensionless
parameters. This classification, which already found a wide
utilization among the chemical combustion community, can be used 
in the field of SNe Ia in order  to emphasize universality in the
physics of turbulent combustion.

In this work we study the properties of turbulent flames in a
certain state, the so-called {\it distributed flame regime}. 
The physical characterization of the latter is given by the
situation where turbulent motions are fast enough to
disturb the flame on microscopic scales. Therefore a distributed
flame does locally not look like a laminar flame anymore which
is a basic difference to turbulent flames in the flamelet regime.
Turbulent burning fronts in an exploding white dwarf are flamelets
at high and intermediate densities, but below $ \sim 5 \times 10^7
\mbox{g cm}^{-3}$ one cannot expect that the flamelet picture
is still valid (\cite{niemwoo97}).

Until now, in the context of SNe Ia, burning in the distributed regime has
only been considered qualitatively. Therefore we wish to present 
first quantitative results regarding this issue. Furthermore, there are good
reasons to believe that this combustion mode plays an
important role in the deflagration to detonation transition (DDT),
which in turn is a promising model, for empirical and theoretical
reasons, for the explosive stage of a SN Ia (e.g. \cite{khok91b};
\cite{niemwoo97}; \cite{niemker97}; \cite{khoketal97}).

In order to attempt a representation turbulent dynamics in the
distributed flame regime, we use a new model, formulated in one
spatial dimension, which nevertheless provides essential features of three
dimensional homogeneous turbulence. It consists of a statistical
description of turbulent mixing and a deterministic evolution of the
underlying microphysics. This method allows a systematical
investigation of turbulence phenomena. When it is coupled to a
nuclear reaction network, it gives first insights of how the flame
structure is affected by turbulence on scales, which have
not been resolved in direct numerical simulations. In
particular, we investigate the flame properties in cases where the
Gibson scale is comparable to the thickness of an undisturbed
conductive flame. The Gibson scale $ l_{\rm G}$ is defined as the
length scale on which
the turbulent velocity fluctuations equal the laminar flame velocity.
In the case of flames in degenerate white dwarf matter $ l_{\rm G}$
becomes comparable to the thickness of the flame only for densities
around $ 2 \times 10^7 \,\mbox{g cm}^{-3}$ and below (\cite{khoketal97};
\cite{niemwoo97}).

The outline of the paper is as follows: first we introduce and describe
the statistical method that we use to model turbulence. Then we couple
the latter to physically relevant microscopical diffusion processes,
such as temperature and viscous diffusion,
and to external energy sources coming from nuclear reactions.
The resulting method is eventually used to investigate some
properties of turbulent flames in the distributed regime, 
like their effective flame speed.
Finally we discuss the relevance of our results to the
DDT problem.

\section{One\--Dimensional Turbulence}

Since fundamental aspects of turbulence can be recovered from the
knowledge of the statistical moments and correlations of the velocity
flow, the statistical approach to turbulence is particulary
appealing. Therefore, the history of statistical methods in turbulence
theory is rather long. In this context we present a novel model of
turbulence (\cite{kers99}). It is a stochastic method, realized as a
Monte Carlo simulation, which allows to compute statistical properties
of the flow velocity and of passive scalars in stationary and decaying
homogeneous turbulence. One-dimensional turbulence (ODT) represents
many aspects of three-dimensional turbulence, but it is
formulated in only one spatial dimension. This model provides the temporal
evolution of a characteristic transverse velocity profile $ u(y,t)$ of
the turbulent medium, where $ y$ is the spatial location on a finite
domain $ [0,Y]$ and $ t$ is the elapsed time. This is done in a
two-fold way: $ u(y,t)$ is subject to a molecular diffusion process
and to a random sequence of profile rearrangements representing
turbulent eddies. Reflecting the typical behavior of turbulence
kinematics, the event rate of these profile rearrangements (so-called
eddy mappings) is proportional to a locally averaged shear of the
velocity profile $ u$.\\ Given the transverse velocity profile at a
certain time, 
$ u(y,t)$, the mapping which models the action of an turbulent eddy of
size $ l$ and at the position $ y_0$ reads
\begin{equation}
	\label{eddymap}
	\hat u(y,t) =  \left\{  \begin{array}{ll}
				\!\!\!u(3y - 2y_0,t)  &  y_0 \leq y \leq y_0 +
	f_1 l\\
				\!\!\!u(-3y +4y_0 + 2l,t) &  y_0 +
	f_1l \leq y \leq y_0 + f_2 l \\
				\!\!\!u(3 y - 2y_0 - 2l,t) &  y_0 +
	f_2 l \leq y \leq y_0 + l 
		        	\end{array}
	       		\right.
\end{equation}
where it is  $ f_1 = 1 - f_2 = \onethird$ and $
  \hat u(y,t) = u(y,t)$ for all $ y \notin [y_0, y_0 + l]$.
This three-valued map represents the typical features of a turbulent
vortex, namely rotation and compression. Each random map defines
an {\it eddy time scale}, $ \tau(y_0,l,t)$, via
\begin{equation}
\	\label{tau}
	\tau(y_0,l,t) = \frac{l}{2\,A\,|u_l(y_0,t) - u_l(y_0 + l/2,t)|} \,\,\,.
\end{equation}
Here, $ u_l$ is the boxcar averaged profile of $ u$ over a length scale $
l$. $ A$ is the only model independent and dimensionless parameter
which has  to be fixed empirically.
Vortical kinetic energy is fed by
the kinetic energy of the local shear. Thus equation (\ref{tau}) can be
interpreted as an energy balance. It then is
\begin{equation}
	\label{energbal}
	\frac{1}{2} \,\rho_0 \,(l\,\tau^{-1})^2 =
	\frac{1}{2} \,\rho_0 \,(2 \, A \, |u_l(y_0,t) - u_l(y_0 +
	l/2,t)|)^2 \,,
\end{equation}
where $ \rho_0$ is the fluid density. Equation (\ref{tau})
resp. (\ref{energbal}) is used to introduce the statistical
hypothesis of ODT. It assumes that the occurrence of eddies with size 
$ l$ and location  $ y_0$ (with respective tolerances $ {\rm d}l$, $
{\rm d}y$) 
is governed by a Poisson random process with mean event rate 
\begin{equation}
	\label{meanrate}
	\frac{A}{l^2 \, \tau(y_0,l,t)} \, {\rm d}y \, {\rm d}l =: \lambda(y_0,l,t)
	\, {\rm d}y \, {\rm d}l \,,
\end{equation}
where the processes for different values of $ y_0$ and $ l$ are
statistically independent. In this context $ A$ can be viewed as a factor
which scales the event rate $ \lambda$. Motivated by the results 
in Kerstein (1999) we choose $A =  0.23$. In addition, the microscopic
evolution is modelled by a diffusion process of the kind $ u_t = \nu
\, u_{yy}$, with the kinematic viscosity $ \nu$. The temporal
evolution of  turbulence, stationary or decaying, depends on the
choice of boundary conditions for $ u$. A decaying turbulent intensity
is obtained by periodic boundary conditions, whereas the stationary
case is given by  the choice of jump periodic boundary conditions: $
u(y+Y,t) = u(y) + u_0, \,\, u_0 := u(Y,t=0)$ and $ u(y,t = 0)$ being
strictly monotonic on $ [0,Y]$.

The numerical implementation of ODT reproduces many typical features
of three-dimensional homogeneous turbulence
(\cite{kers99}). For instance, power spectra of the
kinetic energy show the self similar $ k^{-5/3}$ power law within the
inertial range down to the scale where the transition to dissipation takes
place, c.f. Fig. 1. An obvious advantage of this {\it
ansatz} is the high spatial resolution of turbulence compared with
multidimensional numerical models. In combination with the relatively
moderate computational effort, ODT appears as a useful tool for
performing parameter studies in turbulence theory. On the other hand,
ODT does not consider any pressure fluctuations (dynamical or external) in the
temporal evolution of the velocity profile $ u$. The reason for this
artefact is the inherent conservation of kinetic energy in only one
spatial velocity component due to equation (\ref{energbal}). Note
that the pressure gradient term in the Navier-Stokes equations
redistributes energy among the different spatial components. No such
redistribution is considered here. However, for
isobaric flows ODT appears to be an appropriate model of
turbulence.

\section{Turbulent Flames in Dense C+O Matter}

Although the aim of this work is to investigate some properties
of turbulent burning in dense C+O matter, we begin this section by 
recalling some properties of laminar, conductive
flames. Since the physics of undisturbed flames inside 
a white dwarf is relatively well understood, the first reason
for it is to verify established results, such as those given in
the work of Timmes \& Woosley (1992). The other  reason is that thermal
conduction and viscous diffusion will be employed as the underlying
microscopical processes for ODT in order to model turbulent burning
fronts in white dwarfs.

The state of unburned
matter, i.e. density and nuclear composition, uniquely defines the
propagation velocity $ s_{\rm l}$ of the conductive flame.
Timmes \& Woosley (1992) calculated the flame velocities as well as their
thickness for various fuel compositions and densities.
At lower densities, for $10^8 \, \mbox{g} \, \mbox{cm}^{-3} \gtrsim \rho
\gtrsim 10^7 \, \mbox{g} \, \mbox{cm}^{-3}$, the speed of the laminar 
flame decreases rapidly. This behavior is accompanied by a strong
increase of the flame thickness $ \delta_{\rm l}$, which is
essentially the size of the nuclear reactive zone (\cite{timmwoos92};
\cite{khoketal97}). As already mentioned in the introduction,  we are
especially interested in densities of the order $ \sim (1 - 3) \times
10^7 \,\mbox{g} \, \mbox{cm}^{-3}$. Thus, we set up a
conductive flame propagating into unburned matter consisting of half
$^{12}$C and half $^{16}$O at densities of $ \rho = \, 1.3
\,\times\, 10^{7} \, \mbox{g} \, \mbox{cm}^{-3}$ and of $ \rho = \, 2.3
\,\times\, 10^{7} \, \mbox{g} \, \mbox{cm}^{-3}$. This is done by
solving the equations for the conservation
of mole fractions and enthalpy in planar geometry, viz.
\begin{equation}
\label{abund}
\frac{{\rm d} Y_i}{{\rm d} t} = \sum_{j,k} - Y_i Y_k \lambda_{jk}(i) 
				+ Y_i Y_k
				\lambda_{kj}(i) \,,
\end{equation} 
\begin{equation}
\label{energ}
\frac{{\partial} T}{\partial t} = \frac{1}{\rho c_{\rm p}}\frac{\partial
	}{\partial x} \left( \sigma \frac{\partial T}{\partial x}
	\right) - P \frac{\partial}{\partial t}\frac{1}{\rho}
							+ \frac{1}{c_{\rm p}}\dot{S} \,,
\end{equation}
\begin{equation}
\label{enuc}
\dot{S} = N_{\! A} \sum_{i} \frac{{\rm d}Y_i}{{\rm d}t} B_i \,.
\end{equation}
Herein $ \dot{S}$ denotes the local
specific energy generation rate, $ \sigma $ the thermal conductivity
and $ N_{\! A}$ is Avogadro number and $ B_{i}$ is the nuclear
binding energy of the nucleus considered. The nuclear reaction
network consists of seven species, viz. $^4$He, $^{12}$C,
$^{16}$O, $^{20}$Ne, $^{24}$Mg, $^{28}$Si, and $^{56}$Ni. Equation 
(\ref{abund}) already makes
use of the fact that the Lewis number, that is the ratio of
heat diffusion to mass diffusion, in white dwarf matter is around $
10^7$ and consequently microscopic  transport of the element species can be neglected.
Furthermore, in white dwarf matter the laminar flame speed is much smaller than the speed of
sound, $ s_{\rm l} \,<\, 0.001
a_{\rm s}$. Therefore pressure does not change significantly across the
flame front.

Finally, the typical time scale
for free collapse or free expansion of the star can be estimated by
(\cite{fowler64})
\begin{equation}
\label{hydrot}
\tau_{\rm h} \approx  \frac{446}{\bar{\rho}^{1/2}} \,\,\mbox{s} \,.
\end{equation}
For a white dwarf of radius $R_{\rm wd} = 10^8$ cm and of mass $
M_{\rm wd} = 1.4 \, M_{\odot}$, equation (\ref{hydrot}) gives $
\tau_{\rm h} \approx 0.02$ s.  Timmes \& Woosley
\markcite{timmwoos92}(1992) stated that gravitational influence
can be safely dropped since the diffusion timescale of the flame, 
$ \tau_{d} \approx \delta^2_{\rm l}/D$, where $\delta_{\rm l}$ is the
flame thickness and $ D$ is the
temperature diffusion coefficient, is several orders of magnitude
smaller than $ \tau_{\rm h}$. 
However, for very low densities  ($\rho \lesssim
10^7\,\mbox{g}\,\mbox{cm}^{-3}$) these two
timescales  can become comparable and a possible expansion of the star
could directly affect the laminar flame. To avoid additional
complexity we assume  in this work that the hydrodynamical timescales is
much longer than the relevant timescales of the burning front. 
Thus we consider the small-scale burning front to be not affected by
stellar expansion.

The numerical solution of equations (\ref{abund}) and (\ref{energ}) 
is given explicitly in time combined with an appropriate equation of state
representing ions, black body radiation, electrons and positrons. For
$ \rho = 2.3 \times 10^{7}\,\mbox{g}\,\mbox{cm}^{-3}$, Figure
2  shows the propagation of the laminar flame which is indicated by
the moving jump in the temperature profile. In the steady state
evolution,  the  flame moves into the unburned matter at a velocity $
s_{\rm l} = 2.1 \times 10^{4} \,\mbox{cm s}^{-1}$.  The flame
thickness $ \delta_l$ is characterized by the width of the temperature
jump caused by carbon burning. It is $ \delta_{\rm l} = 1.1$ cm. After
$^{12}$C destruction the burned material has a temperature $ 3.2 \times
\,10^9$ K and a density of $1.45 \times \,10^7 \,\mbox{g cm}^{-3}$.
For these conditions the subsequent oxygen burning has a destruction
timescale  of $ \tau_{\rm des,O} \approx 0.1 $
s. That means that as  the carbon flame propagates with a
steady-state velocity $s_{\rm l}$, oxygen burning ignites at a
distance of $ 2.1 \times 10^3$ cm behind the flame front. This
distance is about ten times larger than our whole computational
domain. Consequently, we are not able to track the whole spatial region
where nuclear reactions take place that are triggered by carbon
burning. The extension of the reactive zone is even larger if one
takes burning of heavier elements, such as silicon, into account. Thus,
the resolution of the flame with all its reactions up to
nuclear statistical equilibrium would require an immense spatial
resolution. 

However, in our case the fuel consists half of
carbon and the destruction of the latter contributes nearly all of the
total energy release by the laminar flame.
In addition, the speed of the
laminar flame is governed by carbon
destruction. Therefore it is fair to concentrate only on the latter.
We also solved the above equations for a density of $ 1.3 \times
10^7 \mbox{ g cm}^{-3}$ with the same nuclear composition as before.
The resulting flame velocity for this case is $ s_{\rm l} = 0.85
\times 10^{4} \mbox{cm s}^{-1}$, while the flame's thickness is 3.8
cm. Our results agree well with the ones given in the work of Timmes \&
Woosley (1992).

As already mentioned in the introduction, the Gibson scale $ l_{\rm
G}$ can be used to measure the influence of turbulence on a
flame. Since it is directly related to the turbulent fluctuation
velocity $ u'$, the Gibson scale depends strongly on the kind of
turbulence that is considered. Essential features of turbulence, like
structure functions, the temporal evolution of the turbulent intensity
and the energy cascade, depend on the absence or presence of certain
geometrical (homogeneity, isotropy) and physical (external forces,
energy sources, state of the turbulent matter) conditions. On large scales in the interior of a white dwarf, turbulence is mainly caused by the
Rayleigh-Taylor instability of the flame which takes place on a length
scale of $ l_{\rm RT} \approx 10^6 \,$cm. This instability produces a
turbulent energy cascade down to dissipative length scales $l_{\rm k}
\approx (10^{-4}-10^{-3}) \,$cm (\cite{khok95}; \cite{niemwoo97}). On
all scales $ l_{\rm k} < l \ll l_{\rm RT}$ turbulence is believed to
be decoupled from gravitational effects and thus can be described by
the Kolmogorov theory. The latter sets the scaling law for $ u'$ to be $
u'(l) \sim l^{1/3}$.  Inbetween these scale bounds nuclear burning is
affected by turbulent motion and different kinds of burning regimes
reveal (For an overview, see \cite{niemwoo97}). Here,
we put our emphasis on the special situation where homogeneous,
isotropic turbulence interacts with the microscopic structure of the
laminar flame, i.e. where $ l_{\rm G}$ becomes comparable to
$\delta_{\rm l}$.  The relation $ l_{\rm G} \approx \delta_{\rm l}$
marks the transition into the distributed burning regime, where the
smallest turbulent vortices can enter the interior of the laminar
flame and can carry away  reactive material before it is completely
burned (\cite{pet86}; \cite{niemwoo97}).  Taking into account the
small Prandtl numbers in the dense matter of a white  dwarf, we can
describe this situation by a constraint on the turbulent Karlovitz
number. This number is the ratio of the diffusion timescale and the 
eddy-turnover-time at the smallest length scale of turbulent heat
transport.
The latter turns out to be the Kolmogorov scale, $ l_{\rm k}$,
only in case of Pr $\approx 1$. For Prandtl numbers much smaller
than one turbulent heat transport is subject to 
significant diffusion effects already at a scale of $ l_{\rm k}
\mbox{Pr}^{-3/4}$. Thus we have
\begin{equation}
\label{karl1}
\mbox{Ka} \equiv \frac{u'(l_{\rm k}) \, \delta_{\rm l}}{s_{\rm l}\, l_{\rm
k}}
\,\mbox{Pr}^{1/2}
 \gtrsim
1 \,.
\end{equation}
Using Kolmogorov theory an
equivalent representation of the latter condition reads
\begin{eqnarray}
\label{karl2}
\mbox{Ka} & = &	\frac{u'(L) \, \delta_{\rm l}}{s_{\rm l}\,
 	L}\,\mbox{Pr}^{1/2}\, \mbox{Re}^{1/2}\nonumber\\
 	  & = & \left(\frac{u'(L)}{s_{\rm l}}\right)^2 
	\, \mbox{Re}^{-1/2} \, \mbox{Pr}^{-1/2}
	 \gtrsim 1 \,.
\end{eqnarray}
% Very similar relations were proposed by Niemeyer \& Kerstein
% (1997)\markcite{niemker97} or by Ronney (1995)\markcite{ron95} 
% for terrestrial conditions. 
There is a
simple relation between the
Gibson length and the Karlovitz number as follows
\begin{equation}
\label{gibson}
l_{\rm G} \approx \delta_{\rm l} \,\mbox{Ka}^{-2} \,.
\end{equation}
Therefore, $ l_{\rm G} < \delta_{\rm l}$ is equivalent to  $ \mbox{Ka}
> 1$. The actual value Ka for a given turbulent
flame plays an outstanding role in turbulent combustion
physics.

We now combine ODT with the structure of a laminar flame to show
how certain properties of a nuclear
flame in dense degenerate matter of white dwarf depend on the Karlovitz number.
The obvious step is to incorporate equation (\ref{energ})
together with a viscous transport evolution for the transversal velocity
profile $ u(x,t)$. Then turbulent advection is modelled by the random eddy
mappings generated by ODT, c.f. equation (\ref{eddymap}). This system reads as
\begin{equation}
\label{odt+diff1}
\frac{{\partial} \hat{u}}{\partial t} =  \frac{\mathrm{
Pr}}{\hat{\rho} \,c_{\rm p}}\frac{\partial
	}{\partial x} \left( \sigma \frac{\partial (\hat{u})}{\partial x}
	\right) \,,
\end{equation}
\begin{eqnarray}
\label{odt+diff2}
\nonumber
\frac{{\partial} \hat{T}}{\partial t} & = & \frac{1}{\hat{\rho}\,c_{\rm p}}\frac{\partial
	}{\partial x} \left( \sigma \frac{\partial \hat{T}}{\partial x}
	\right) - P \frac{\partial 1/\hat{\rho}}{\partial t}
							+
	\frac{1}{c_{\rm p}}\dot{S} \,,
\end{eqnarray}
where $ \hat{u}, \hat{\rho}, \hat{T}$ are the rearranged profiles of
the velocity, the density and the temperature according 
to the eddy-mapping, c.f. equation (\ref{eddymap}).

We choose the amplitude of velocity fluctuations of $ u$ to be $\sim
10^7$ cm s$^{-1}$ at $ \sim 10^6$ cm, and to obey Kolmogorov scaling.
These values are motivated through the expected speed of buoyant unstable hot
bubbles of size $ L \approx 10^6$ cm, which become Rayleigh-Taylor
unstable and eventually  give the main contribution to turbulent energy on large scales.
 Having fixed this velocity at a 
certain lengthscale, we can use Kolmogorov scaling to estimate
 $ u'(l)$ on smaller scales $ l$. 
Thus the amplitude of turbulent velocity fluctuations in our model is 
comparable to the expected small-scale velocity fluctuations within
a SN Ia. 

However, due to computational limitations we cannot consider $ u$ to
be the actual fluid velocity of the white dwarf matter. Since its
Prandtl number is around Pr $\gtrsim 10^{-5}$, the numerical resolution of
both, the viscous and the heat conductive diffusion scale, would cost
too much numerical effort even in one spatial dimension. To get out
of this dilemma, we use an 'artificial' Prandtl of $2
\times 10^{-2}$ and of $ 10^{-1}$. This choice
underestimates the inertial range of turbulent motions within
an exploding Chandrasekhar-mass white dwarf. But we believe that the physics of distributed burning in this kind of matter is decoupled from scales much smaller
than $ \l_{\rm k} \,\mbox{Pr}^{-3/4}$, because turbulent eddies of sizes
well below this limit will be smeared instantly out by temperature
diffusion. Therefore, we do not essentially change the flame properties
in the distributed burning regime by increasing Pr. Of course, this is
only valid for sub-unity Prandtl numbers, otherwise one possibly
leaves the distributed regime.

%\vfill
The random velocity field $ u$ is generated by stationary ODT, thus $ u'$
becomes constant in time after an initial transient phase.
 For our studies we perform simulations where  $ u'$ (at a length
scale $ l = 1$ cm) is equal or larger than a given value $s_{\rm
l}$. Then we estimate some flame properties
such as the effective turbulent flame
speed, $ s_{\rm T}$. So far, analytical
results do not predict the function $ s_{\rm T}(u'; \mbox{Ka})$
correctly. For instance, experimentally observed {\it bending
effects} or even complete flame quenching are still mostly unexplained
problems (Ronney 1995).

%\vfill
An overview of the various numerical models is shown in 
Table \ref{tbl-1}. Each realization is uniquely parametrized by its 
Reynolds, Karlovitz and Prandtl number.
To get the turbulent Reynolds numbers, Re $= u'(L)\, L/\nu$,  we
have estimated the integral scale $ L$ for each model from the
wavenumber $ k_L$ for which the relevant energy power spectrum $ E(k)$
has a maximum.  Together with the r.m.s. turbulent velocity
and the kinematic viscosity, $\nu = {\rm Pr} (\rho c_{\rm p})^{-1}
\sigma$, we eventually obtain  the Reynolds number Re and the Karlovitz number Ka.
An example of a numerical realization is 
illustrated in Figure 3, where a laminar and a turbulent flame
are shown in a space-time diagram. 

Table \ref{tbl-1} also shows the measured turbulent flame speeds
$s_{\rm T}$ as a function of Ka, and Pr. These velocities have
been obtained by measuring the slope of the function 
\begin{equation}
\Delta(t) = \int c(x,t) \,{\rm d}x \,, 
\end{equation} 
where $c(x,t)$ is the reaction-progress variable defined by (the indices b 
and u assign the burned and unburned states)
\begin{equation}
c(x,t) = \left( \frac{T(x,t)}{T_{\rm u}} - 1 \right) \,\frac{T_{\rm
u}}{T_{\rm b} - T_{\rm u}} \,.
\end{equation}
Thus $ s_{\rm T} = {\rm d} \Delta(t)/ {\rm d}t$ defines the effective
flame speed. This value is constant only in the
case of laminar flames. However, in the turbulent case it is 
subject to fluctuations. 
These estimates enable us to express the dimensionless turbulent
flame speed, $ U_{\rm T} = s_{\rm T}/s_{\rm l}$, as a function 
of the dimensionless turbulent fluctuation velocity, $ U = u'/s_{\rm
l}$. We should mention, that we computed the dimensionless turbulent
flame speed using the alternative formula
\begin{equation}
\label{utniem}
 U_{\rm T} = \int \rho_{\rm t}
 \dot{S}_{\rm t}(x) \,{\rm d}x \,(\!\int \rho_{\rm l}(x) \dot{S}_{\rm
 l}(x) \,{\rm d}x )^{-1} \,,
\end{equation}
where $ \dot{S}_{\rm t}$ and $ \dot{S}_{\rm l}$ denotes the
turbulent and laminar profile of the energy generation rate. 
In fact, the estimated values 
are  in good agreement with the ones obtained through the equation 
$ s_{\rm T} = {\rm d} \Delta(t)/ {\rm d}t$. 

For our studies we vary the r.m.s. turbulent velocity,
the Prandtl number and the density while the integral length scale $ L$
remains at a constant value.
%\vfill

The results are shown in Figure 4. It clearly demonstrates that
the turbulent flame speed at high turbulent strain rates, $ U \gg 1$,
does not follow a law valid in the flamelet regime. At a Karlovitz 
number greater than one, i.e. in the distributed burning regime,
there is a significant deviation from this law: stronger turbulent
intensities do not necessarily lead to a faster burning rate. 
This effect is more pronounced in the low-density case, where
higher turbulent velocities even cause a slight deceleration of the
flame speed.   
This result confirms the 
intuitive picture that fast flames (flames at higher density)
are more resistant to distortion through turbulence than
the slower ones (flames at lower density).
%\vfill

As already mentioned in this section, the transport of heat
is expected to be decoupled from turbulence on scales smaller than 
$ l_{\rm k} \mbox{Pr}^{-3/4}$. This assertion is numerically
confirmed in the realization B1, where we significantly decrease
the Kolmogorov length scale (using a Prandtl number of $ 0.02$)
while keeping the fluctuation velocity $ u'(L)$ at a value used
in the realization A7 (Pr = 0.1). The estimated turbulent 
flame speed $ s_{\rm T}$ of both models however turns out to be nearly the
same, see Figure 4. This behavior suggests that in cases of a small
Prandtl number one  should use the {\it Peclet} number,
$\mbox{Pe}  = \mbox{Re Pr}$, as the governing parameter 
for the strength of turbulent flame dynamics.

For clearity, Table 2 gives a summary of 
all the dimensionless quantities that we have used or mentioned 
in this section.

\section{Discussion}
\subsection{Burning in the Distributed Regime}

The coupling of ODT as a model of homogeneous turbulence to reactive
hydrodynamics gives an insight into the physics of distributed
burning. First of all the qualitative physical picture of this burning
process is revealed: turbulent vortices enter the reaction zone and
carry away burning material to the front and to the back of the
flame. As a consequence, one observes a completely different behavior than in the flamelet
regime, because unlike the latter, the distributed regime 
changes the local flame velocity (The local flame velocity is 
the speed at which the flame propagates locally normal to itself.).

A different feature is the existence of localized burning with
high burning rates, commonly referred to as a local explosions
(\cite{radchan95}). Turbulent motion forms regions where hot reaction
products are mixed with cold unburned material. These regions have a
higher temperature than pure fuel and possess comparably short
destruction timescales for carbon.  Their eventual re-ignition causes
high burning rates, as shown in Figure 5. We find
that with higher Karlovitz number more of these events occur and
their strength (burning rate) is increased. Terrestrial experiments
show the generation of blast waves after the occurrence of local
explosions.  In some experiments the strongest waves developed even
into a detonation (\cite{radchan95}). However, laboratory experiments
on turbulent flames deal with much higher expansion factors,
$\rho_{\rm u}/\rho_{\rm b}$, than those given by flames in white dwarf matter. Thus we expect the effect
of blast waves in the context of SNe Ia to be less effective than in typical
chemical flames.
However, because our model does not consider pressure waves,
the whole effect of local explosions on the burning process cannot be
presented here and we cannot completely exclude the possibility that
blast waves 
possibly enhance the global burning rate by compression and
pre-heating of the surrounding medium. Neglecting such effects in
our model possibly means an underestimation  of
turbulent combustion rate in the  distributed regime. 

Taking into account the estimates on the turbulent flame speed of Table \ref{tbl-1}
we obtain a relatively good 
agreement with experimental data presented by Abdel--Gayed et
al. (1987), henceforth AG97. Their
work contains about 1650 experimental results on turbulent flame velocities
at various strain rates $ U = u'/s_{\rm l}$, fuels and equivalence
ratios. Their results were obtained under conditions far from
those in an exploding Chandrasekhar-mass white dwarf, but it is worth to
verify if certain physical statements, like the function $ U_{\rm
T}({\rm Ka}; {\rm Re})$, appear to show universality. These experimental
results and earlier work (\cite{abdel85}) present two
important phenomena.  First, for a fixed Reynolds number, $U_{\rm T}$
is not a strong monotonic function of the strain rate $ U$. Rather, it
reaches a maximum at a certain turbulent intensity and eventually
saturates or even decreases for higher values of the strain rate
(bending effect). As a second feature, there is flame
extinction at a Karlovitz number Ka $\gtrsim 7$ for Reynolds numbers
Re $> 10^2$. Figure 6 shows several
experimemtal values of the function $ U_{\rm T}$ taken from the work of AG87.
These values correspond to the same sample of ratios $ u'/s_{\rm l}$ we
use in our numerical caluclations.

Our numerical results do show the bending
effect for high strain rates which is in agreement with laboratory
experiments. However, they do not
show complete flame quenching, i.e. a situation where fuel
is not consumed anymore. This is not surprising because the basic laws
of thermodynamics prevent a turbulent (or
laminar)  flame from complete quenching. The first law of
thermodynamics states that the reactive mixture {\it must} eventually
reach the adiabatic flame temperature. The second law demands that
heat {\it must} be transported from the burned to the unburned
regions. Therefore, there is no way to prevent consumption of
unburned material in a system without possible heat losses. 
And clearly, unlike under laboratory conditions, the flame 
in an exploding white dwarf is not subject to significant 
heat losses.
%vfill

The comparison with our results 
shows that the flame speeds presented here are lower than those in AG87.
 We think that one reason for this discrepancy is the
different value of the Lewis number in our work and in the data of
AG87. It is known that high Lewis number flames are more sensitive
to strain than flames with a low value of Le. In fact, analytical
results on planar laminar flames show that the flame speed 
decreases when a velocity field with strain, i.e. a non-vanishing
gradient of this velocity field, is invoked parallel to 
the flame surface. One finds that at higher
Lewis numbers the decrease of the flame speed is stronger (\cite{tromans81}).
Therefore we expect thermonuclear flames in C+O matter to be more
sensitive on strain effects, because
of their Lewis numbers of aprrox. $ 10^7$.
Another source of this deviation could be the already aforementioned
difference in the expansion factors between laboratory and white dwarf
combustion.

Thermonuclear flames in the distributed regime thus experience
a two-fold effect through turbulence: One is the acceleration
due to  an enhancement of the heat transport 
produced by small-scale turbulence. The other is a slowing-down
due to additional turbulent strain, which can be measured by the
Karlovitz number. However, we have to stress that we do not have 
revealed the actual physical mechanism responsible for the
deceleration. This is the task of future work.

\subsection{Implications for a Deflagration to Detonation Transition}

As already stated by many authors, a delayed detonation within the explosion
of a Chandrasekhar-mass white dwarf  is a reasonable model. It is a
good fit to experimental constraints for element abundances obtained from
observed spectra. In addition, although it is almost clear that after
ignition the burning process evolves as a deflagration, there is no
fundamental physical objection why a transition to detonation in the
subsequent stages of the explosion should be impossible. Therefore, DDT
models should be taken into account in supernova theory. In the
context of this work, we discuss the possibility of distributed burning  
leading to a detonation.

Based on the Zeldovich gradient mechanism, Khokhlov et al. (1997)
calculated the minimum size, $ L_c$, of a region where fuel and ashes
are mixed and out of which a stable detonation wave can emerge. In
their model the formation of this region is caused by turbulent
mixing, but it is not obvious how this process is realized. They
 proposed that global expansion of the star along with the following
density drop quenches the deflagration flame, whereas the
Rayleigh-Taylor instability still takes place on scales of
$10^6$--$10^8$ cm. Before the star recompresses again, fuel and ashes
are mixed and turbulence eventually produces a complete mixed
region of size of the order of $ L_c$. Thus, pulsation may lead to a DDT.

Apart from considering pulsations of the star, it is worth asking
whether burning in the distributed regime alone can lead to a
DDT. Niemeyer \& Woosley (1997)  advocated this burning mode as a
favorite model for DDT.  Basically, they proposed that if turbulent
burning is able to form a region of uncompletely burned material with
a certain temperature gradient, then a carbon detonation can occur
when two presumtions are fulfilled.  First, the aforementioned region
must have at least a size of the critical radius necessary for a
detonation (These critical radii could be calculated explicitly for
different densities and nuclear compositions, see Niemeyer \& Woosley
1997.).  Second, the temperature gradient must be shallow enough in
order to allow a sonic phase velocity of nuclear burning.  For
instance, consider a mixed region consisting of half carbon and
oxygen at a density of $ 3 \times 10^7 \mbox{g cm}^{-3}$ and having
a maximum temperature of $ T_{\rm max} = 2 \times 10^9$ K.  It then
follows that this region must have a size of at least $ L_c \sim 50$
m.  Furthermore, temperature differences within it should nott exceed 
$ \sim 10^5$ K. Then a stable detonation wave can form. 

Both approaches have in common that they require  a rather shallow
temperature gradient over a region with a sufficient amount of
carbon. Until now, it has  not been demonstrated how turbulence can
 achieve or lead to such a high level of temperature homogeneity.

In this context it is interesting to know
the maximum length scale, $ l_{\rm max}$, where distributed burning in
a white dwarf can occur.
A lower bound of this scale is given
by the size of those eddies whose turnover time equals a typical
turbulent burning timescale $ \tau_{\rm nuc}$. That is 
\begin{equation}
\label{lmax1}
  l_{\rm max}/u'(l_{\rm max}) = \tau_{\rm nuc} \,.
\end{equation}
In the case of laminar flames $
\tau_{\rm nuc}$ is just the diffusion timescale $ \tau_{\rm l} \approx
\delta_{\rm l}/s_{\rm l} \approx 0.7 \times 10^{-4}$ s.  But what is
the value $ \tau_{\rm nuc}$ in the presence of turbulence?  We argue
that even in the turbulent case the choice of $ \tau_{\rm l}$ as
the {\it typical} burning timescale is natural. To justify this assumption, we
compute the spatial distribution of destruction timescales across the
flame. In general, this distribution is characterized by a minimum value $
\tau^*_{\rm nuc}$. It corresponds to the location of the maximum energy
generation within the flame. In our numerical models, we find that $
\tau^*_{\rm nuc}$ is independent of the turbulent intensity, i.e. its
value is the same for the laminar case and for the distributed regime.
In fact, $\tau^*_{\rm nuc}$ equals $ \tau_{\rm l}$.  
Then equation (\ref{lmax1}) leads to  the useful expression
\begin{equation}
\label{lmax}
l_{\rm max} \approx \delta_{\rm l} \, \mbox{Ka} \,.
\end{equation}
This number can be viewed as the size of the distributed flame, i.e.
an estimate on the size of the interface that seperates burned
and unburned material. However, there are reasons
to believe that in reality $ l_{\rm max}$ exceeds the value
given by equation (\ref{lmax}). This is because there are turbulent vortices
with bigger size than $ l_{\rm max}$ that can enter the flame
and transport burning material with longer burning times than
$ \tau_{\rm nuc}$.
Hence, we do justice to these considerations and 
set the effective value of $ l_{\rm max}$ to
\begin{equation}
\label{lmaxalfa}
l_{\rm max} \approx \alpha \,\delta_{\rm l} \, \mbox{Ka} \,,
\end{equation}
with $ \alpha > 1$.
The actual value of $ \alpha$ however, i.e. the ability of turbulence to 
stretch out the flame brush, has to be estimated by
more rigorous and quantitive considerations, see Lisewski et al. (1999). 

If $ l_{\rm max}$ itself
reached the critical size $ L_c$ required for a detonation, then
burning in the distributed regime would become more of a promising model for
DDT. But still it would not be clear if present temperature
fluctuations would be small enough to allow this transition to happen. 
Thus the equality of $ l_{\rm max}$ and $ L_c$ is only a
necessary condition for a DDT. At a density of $ 2.3 \times 10^7 \mbox{g
cm}^{-3}$, Khokhlov et al. (1997) obtain $ L_c \approx 10^4$ cm.
Now the expected turbulent velocities at an integral scale of
$ L = 10^6$ cm are $ u' \sim \sqrt{g_{\rm eff} L} \sim
10^7 \mbox{cm s}^{-1}$, where $ g_{\rm eff} \sim 10^8 \mbox{cm
s}^{-2}$ is the effective gravitational acceleration. Hence, at
densities $\rho \gtrsim 2 \times 10^7 \mbox{g cm}^{-3}$ the
corresponding  Karlovitz numbers are only around Ka $ \approx 10$. 
According to equation (\ref{lmax})  a very high Karlovitz number Ka $
\approx 10^4$ is required to obtain equality of $ l_{\rm max}$ and $
L_c$.
Consequently, the enhancement-factor $ \alpha$ in equation (\ref
{lmaxalfa}) would have to be of the order of $ 10^3$. This means that
the ashes-fuel interface has  to be extended much more in size than
given by equation (\ref{lmax}).\\
Also at a density of $ 10^7 \mbox{g cm}^{-3}$,
where $ s_{\rm l} = 4.7 \times 10^3 \mbox{cm s}^{-1}$ and $
\delta_{\rm l} = 4.2$ cm (Timmes \& Woosley 1992), the value of $
\alpha$ has still to be around $ 10^3$ in order to enable a direct DDT.

Furthermore, the process of a DDT demands, that the mixing timescale, $\tau_{\rm
mix}$, which can be regarded as the time it takes turbulence to form
a complete mixed region of size $ L_c$, must be smaller than a typical
burning timescale, $ \tau_{\rm nuc}$, of the mixed turbulent matter. But in
order to avoid quenching by expansion, $\tau_{\rm nuc}$ in turn must be
smaller than the hydrodynamical time of the star, $ \tau_{\rm h}$,
c.f. equation (\ref{hydrot}). These conditions read as $
\tau_{\rm mix}\,  <   \tau_{\rm nuc}  <  \, \tau_{\rm h}$, or as
\begin{equation}
\label{chaintau}
	\frac{L_c}{u'(L_c)} < \tau_{\rm nuc}(l_{\rm max} = L_c)  
	<  \frac{446}{\bar{\rho}^{1/2}} \,.
\end{equation}
For $ \rho = 2 \times 10^7 \mbox{g cm}^{-3}$, we have $ \tau_{\rm mix} \approx
5$ ms while the hydrodynamic time is $ \tau_{\rm h} = 0.02$ s.
Here now the question arises whether turbulence can form such
a region whose burning time is smaller than $ \tau_{\rm h}$.
This problem is adressed to forthcoming work, where we study -- by
using ODT -- the burning times on different lengthscales within the distributed flame.
However, until we do not have a better understanding of the large scale
structure (lenght-scales of $ \gtrsim L_c$) of flames in the distributed regime, the preliminary result
from equation (\ref{lmax}) suggests
that Karlovitz numbers in white dwarf C+O matter at densities around $ 10^7 \mbox{g
cm}^{-3}$ are too
small to form a region of critical size necessary for a DDT. 

\section{Conclusions}
	
The aim of this work was to investigate the physics of distributed
burning in an exploding white dwarf of chandrasekhar mass. With ODT as
a new model for homegeneous, isotropic turbulence
we are able to study the effective turbulent flame speeds 
resulting from  this kind of burning in dense C+O matter. Our results show that in this regime the local
properties of the thermonuclear flame are readily changed:
the flame cannot be represented by the flamelet model anymore 
and moreover, higher turbulent intensities in general can lead to 
a lower local flame speed. This behaviour has already been observed
in laboratory experiments and is introduced the first time
in the context of nuclear explosions in type Ia SNe.
We have also found that, in cases where the
Gibson length $l_{\rm G}$ is smaller than $ \delta_{\rm l}$
and where the Prandtl number is much smaller than one, turbulent burning  
decouples from turbulent scales smaller than $ l_{\rm k} \mbox{Pr}^{-3/4}$.
This suggests that the appropriate parameter for turbulent heat
transport inside
a white dwarf is given by the Peclet number instead of the Reynolds number. 
Because of the small ratio of viscous momentum to diffusive heat transport
the former is by several orders of magnitude
smaller than the Reynolds number: $ \mbox{Pe} = \mbox{Pr Re}$.

In addition, first quantitative considerations
demand that a turbulence-induced deflagration to detonation transition can only
happen deep in the distributed burning regime, i.e. at Karlovitz numbers
larger than $10^3$. However, our estimates for the turbulent velocity at
the integral scale in  an exploding white dwarf give values of Ka that
are too small (at least by a factor of $10^3$) in order to make a direct DDT probable. 

Finally, it is wortwhile to remark that we have completely left out questions about the influence of
the turbulent spectrum on this issue.  We have only shown that a DDT
is unlikely, when  Kolmogorov theory of turbulence is
considered. Niemeyer \& Kerstein (1997) proposed that turbulence in an
exploding white dwarf is dominated by a {\it potential} energy cascade
instead of Kolmogorov's kinetic energy cascade. This  domination leads
to different scaling properties (Bolgiano-Obhukov scaling) of the
turbulent velocity. As a consequence, higher turbulent velocities are
necessary for the transition to distributed burning. Furthermore,
Niemeyer \& Woosley give the speed of the  turbulent fluctuations, $
v_{\rm RT} \approx 0.5 \sqrt{g_{\rm eff} L} \approx u'(L)$, which may
not be the upper limit for turbulent velocities at the largest
turbulent scale. It is based on the original idea of Kerstein (1996),
that the turbulent flame velocity, $ s_{\rm T}$ is increased by local
expansion caused by thermonuclear reactions ({\it active turbulent
combustion}). At the largest length scale, $ L$, and at a certain
temporal stage of the burning process, the speed of the flame may
exceed $ v_{\rm RT}$. Because the relation $ u'(L) \approx s_{\rm T}$
still holds at this point, a significant enhancement of turbulent
velocity fluctuations may occur. But it is unclear, if the growth
rate, $ \dot{s}_{\rm T}/s_{\rm T}$, is high enough to realize this
enhancement within a hydrodynamical time $ \tau_{\rm h}$.\\ These
arguments favour a DDT, because their effect is an increase of the
turbulent fluctuations on lengths comparable to $ L_c$.  However, both
arguments are rather speculative, because they have not been confirmed
numerically, analytically or experimentally in the context of SNe Ia
yet.
\medskip

This work was supported by the Deutsche Forschungsgemeinschaft and
the Deutscher Akademischer Austauschdienst.
AML kindly acknowledges partial support by the NSF grants
INT-9726315 and AST-3731569.
Support from the Office of Basic Energy Sciences, U.S. Department of 
Energy, is also acknowledged (ARK).

%
%-----------------------
%

%--------------------------
%table 1
%--------------------------
\begin{deluxetable}{ccccc}
\footnotesize
\tablecaption{Realized models of turbulent flames.
 X($^{12}$C) = X($^{16}$O) $= \onehalf$,
 $\rho = 2.3 \times
10^7\,\mbox{\rm g cm}^{-3}$ (A1 -- A9, B1), $\rho = 1.3 \times
10^7\,\mbox{\rm g cm}^{-3}$ (C1 -- C5, C41, C51).
\label{tbl-1}}
\tablewidth{0pt}
\tablehead{
\colhead{Model} & \colhead{Reynolds Number}   & \colhead{Karlovitz
Number}	& \colhead{Prandtl Number}  & \colhead{Flame Speed}\nl
\colhead{} & \colhead{Re}   & \colhead{Ka}  & \colhead{Pr}
 &\colhead{$s_{\rm T}$ [$\times 10^4 $ cm s$^{-1}$]}
}
\startdata
A1 &114  &0.06  &0.1  &$2.15 $\nl
A2 &270  &0.23  &0.1  &$2.4 $\nl
A3 &452  &0.51  &0.1  &$2.7 $\nl
A4 &598  &0.78  &0.1  &$3.1 $\nl
A5 &673  &0.92  &0.1  &$3.4 $\nl
A6 &820  &1.24  &0.1  &$4.2 $\nl
A7 &1285  &2.43  &0.1  &$5.0 $\nl
A8 &1493  &3.05  &0.1  &$5.3 $\\[.2cm]
B1 &6182  &2.29  &0.02  &$4.9 $\\[.2 cm]
C1 &28  &0.68  &0.1  &$1.0 $\nl
C2 &119  &5.80  &0.1  &$1.8 $\nl
C3 &212  &13.9  &0.1  &$2.4 $\nl
C4 &314  &25.7  &0.1  &$2.1 $\nl
C41 &323  &25.3  &0.1  &$2.2 $\nl
C5 &366  &32.4  &0.1  &$1.9 $\nl
C51 &371  &32.2  &0.1  &$2.0 $\nl
\enddata
\end{deluxetable}
%--------------------------
%table 2
%--------------------------
\begin{deluxetable}{clcc}
%\footnotesize
\tablecaption{Summary of some important dimensionless parameters.
\label{tbl-2}}
\tablewidth{0pt}
\tablehead{
\colhead{Symbol} & \colhead{Name}   & \colhead{Mathematical
definition}	
& \colhead{Physical description}
}
\startdata
{\rm Re} &Reynolds number  &$\displaystyle{\frac{u'(L) \,L}{\nu}}$  &$\displaystyle{\frac{\mbox{integral scale
turbulent diffusivity}}{\mbox{viscous diffusivity}}}$\\[.4 cm]
{\rm Pr} &Prandtl number &$\displaystyle{\frac{\rho c_{\rm p}
\nu}{\sigma}}$  
&$\displaystyle{\frac{\mbox{viscous diffusivity}}{\mbox{heat
diffusivity}}}$\\[.4 cm]
{\rm Pe} &Peclet number  &Re Pr  &$\displaystyle{\frac{\mbox{integral scale
turbulent diffusivity}}{\mbox{heat diffusivity}}}$\\[.4 cm]
{\rm Ka} &Karlovitz number  &$
\displaystyle{\frac{u'(l_{\rm k}) \,\delta_l}{s_{\rm l} l_{\rm k}} 
 \mbox{{\rm Pr}}^{1/2}}$
&$\displaystyle{\frac{\mbox{heat diffusion timescale}}{\mbox{smallest
turbulent heat transport timescale}}}$\\[.4 cm]
{\rm Le} &Lewis number  &$\displaystyle{
\frac{\sigma}{\rho c_{\rm p} D_{\rm m}}}$
&$\displaystyle{\frac{\mbox{heat diffusivity}}{\mbox{mass
diffusivity}}}\,\,$\tablenotemark{\dagger}\\[.4 cm]

\enddata
\tablenotetext{\dagger}{$D_{\rm m}$ denotes the ass diffusion coefficient.}
\end{deluxetable}
%
%--------------------------
%
%--------------------------
%eps
%--------------------------
\begin{figure*}[ht]
\figurenum{1}
\plotone{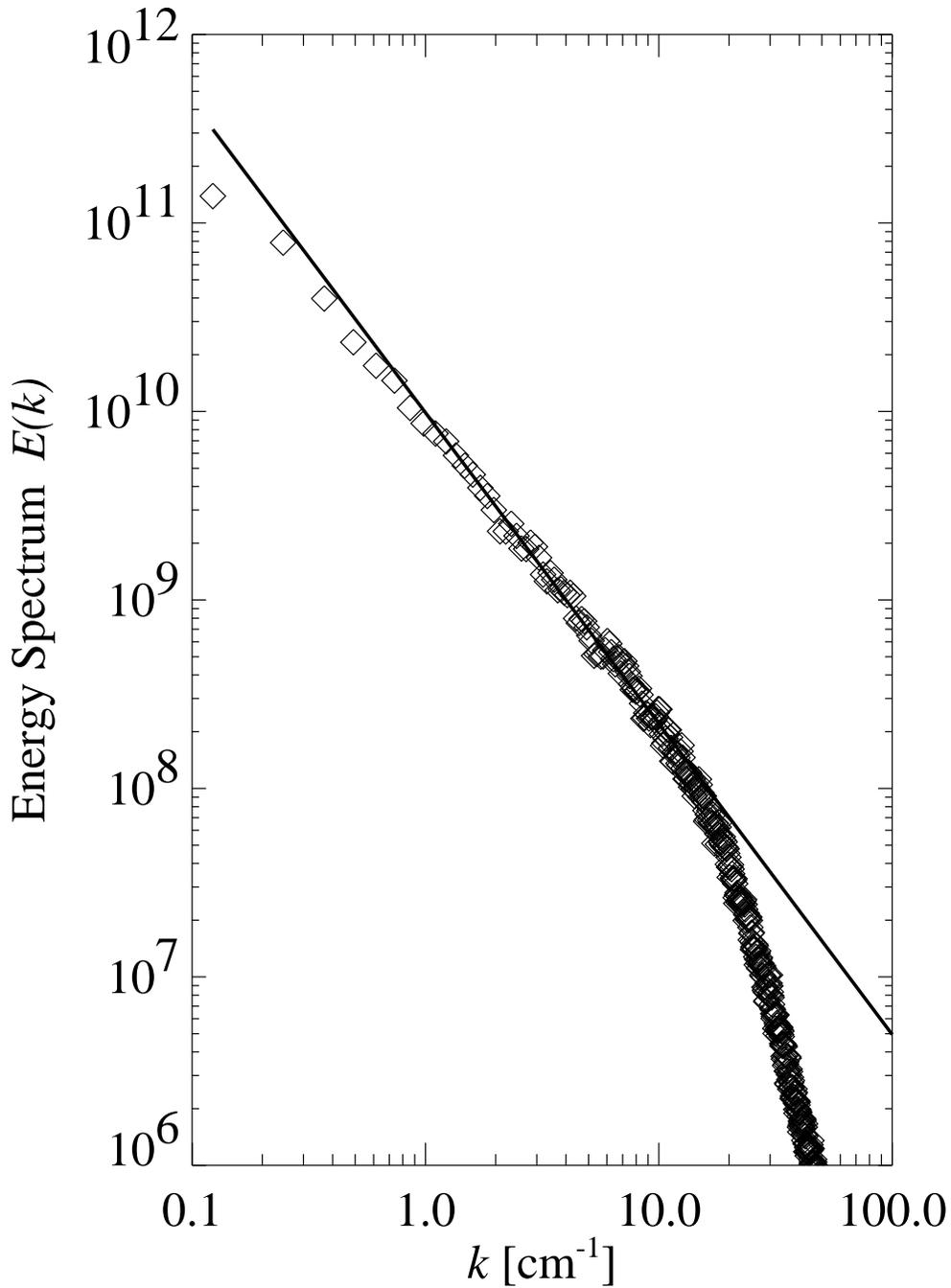}
\caption{An energy power spectrum generated by
stationary ODT ($ \diamond$). Up to a constant factor the power
spectrum is the Fourier transform of the second structure function $
S_2 = \langle(u(r) - u(0))^2\rangle$. The solid line represents the
$k^{-5/3}$ law for the inertial range predicted by Kolmogorov
theory.}
\end{figure*}
%
%--------------------------
%eps
%--------------------------
\begin{figure*}[ht]
\figurenum{2}
\plotone{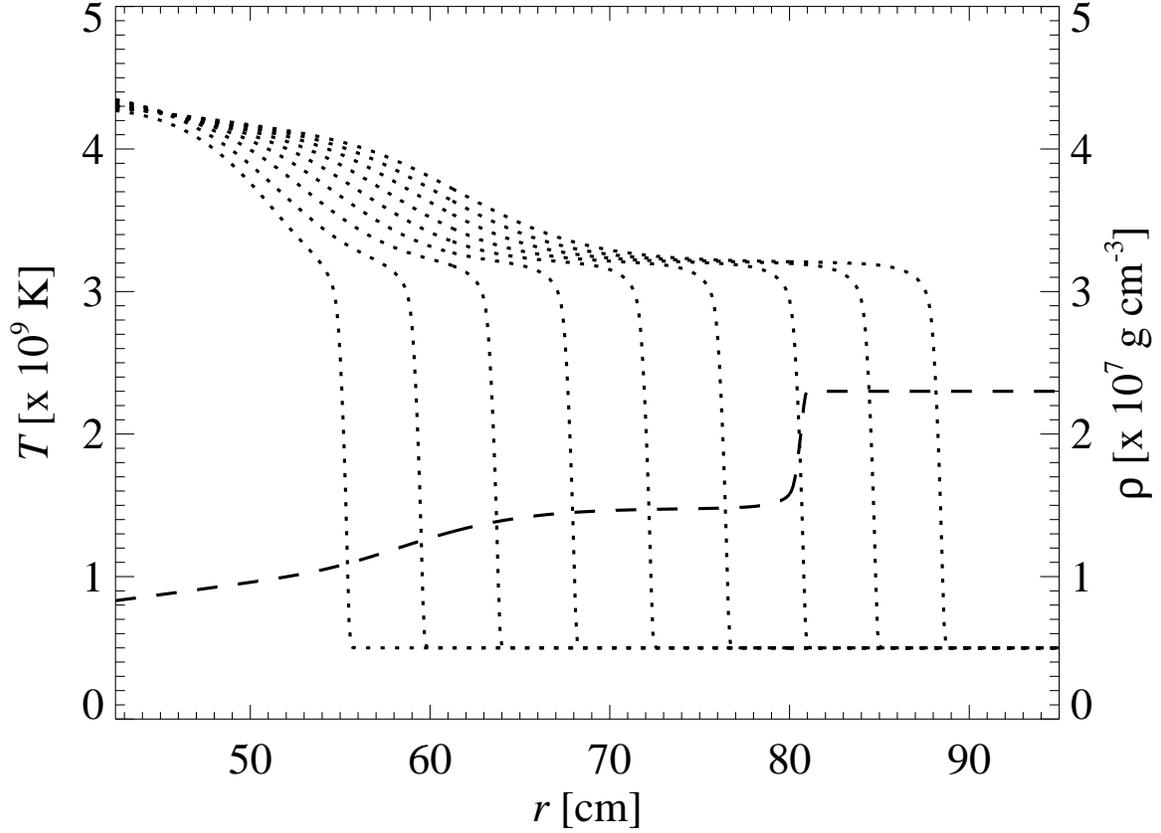}
\caption{The propagation of a X($^{12}$C) = X($^{16}$O) = $0.5$
conductive flame. The sequential temperature profiles (dotted lines)
are shown at times $t =  1.3 \,\times\, 10^{-4}, \,3.0 \,\times\,
10^{-4},\, 6.5 \,\times\, 10^{-4}, \, 8.3 \,\times\, 10^{-4}, \, 1.01
\,\times\, 10^{-3},\, 1.19 \,\times\, 10^{-3}, \,1.36 \,\times\,
10^{-3}, \, 1.51 \,\times\, 10^{-3}$ s. The dashed line is the density
profile at $ t = 1.19 \,\times\, 10^{-3}$ s. The carbon flame already
moves in a steady-state, while the temperature rise at the
left-hand-side comes from oxygen burning due to the energy stored
at the initial state of the simulation.}
\end{figure*}
%--------------------------
%eps
%--------------------------
\begin{figure*}[ht]
\figurenum{3}
\plottwo{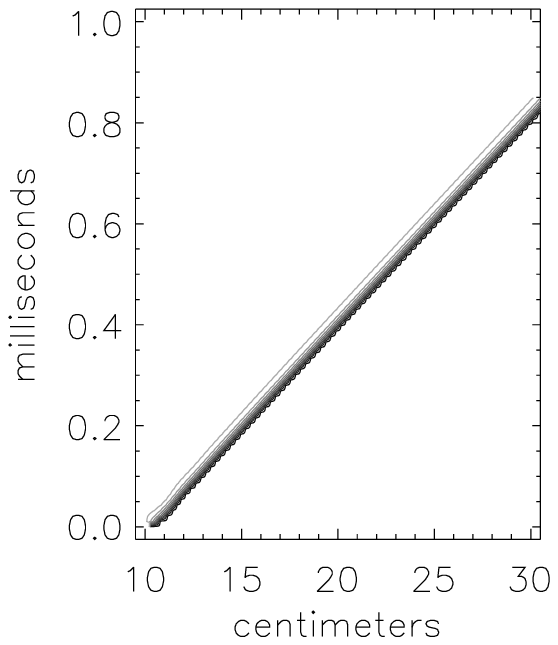}{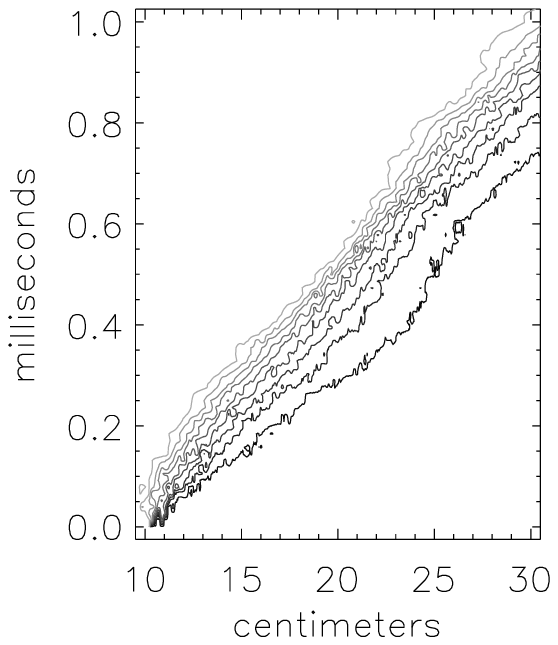}
\caption{Comparison of a laminar flame (left panel) propagating into C+O matter
with a turbulent one (right panel) generated with ODT. The lines represent
iso-temperature regions at values from $0.6$ to $3.0 \times 10^9$ K. The right panel shows data from the numerical realization A2.}
\end{figure*}
%--------------------------
%eps
%--------------------------
\begin{figure*}[ht]
\figurenum{4}
\plotone{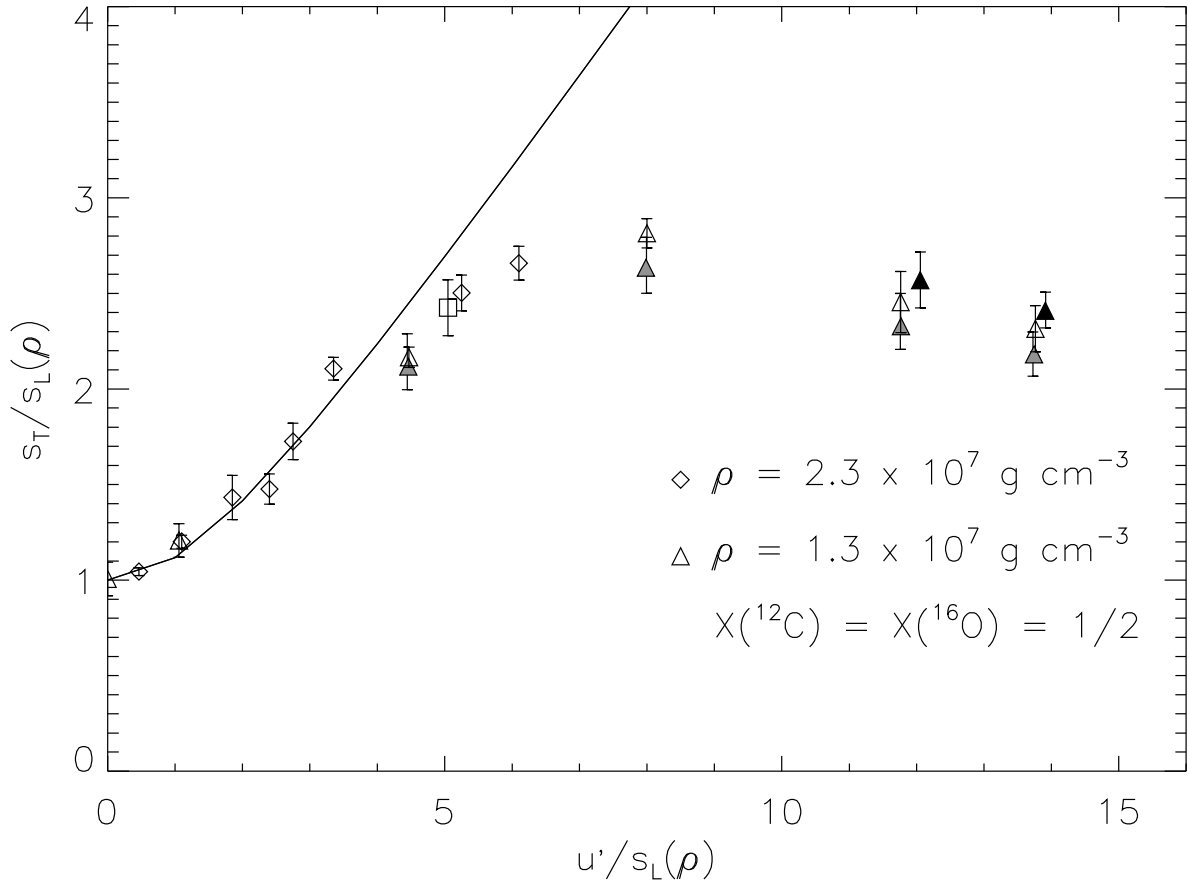}
\caption{The dimensionless turbulent flame speed as a function of
the dimensionless r.m.s. turbulent velocity fluctuations. At high 
strain rates there is a strong deviation from 
the flamelet model, which predicts $ U_{\rm T} = (1 + \beta
U^2)^{1/2}$ (solid curve with $ \beta = 0.24$), (see for instance \cite{poch94}).
The numerical realization B1 is depicted by a box ($ \Box$). The two black
triangles give the results from the realizations  C41 and C51 having
a four-times higher spatial resolution than C4 and C5. The gray 
triangles represent those speeds obtained from formula (\ref{utniem}).}
\end{figure*}
%--------------------------
%eps
%--------------------------
\begin{figure*}[ht]
\figurenum{5}
\plotone{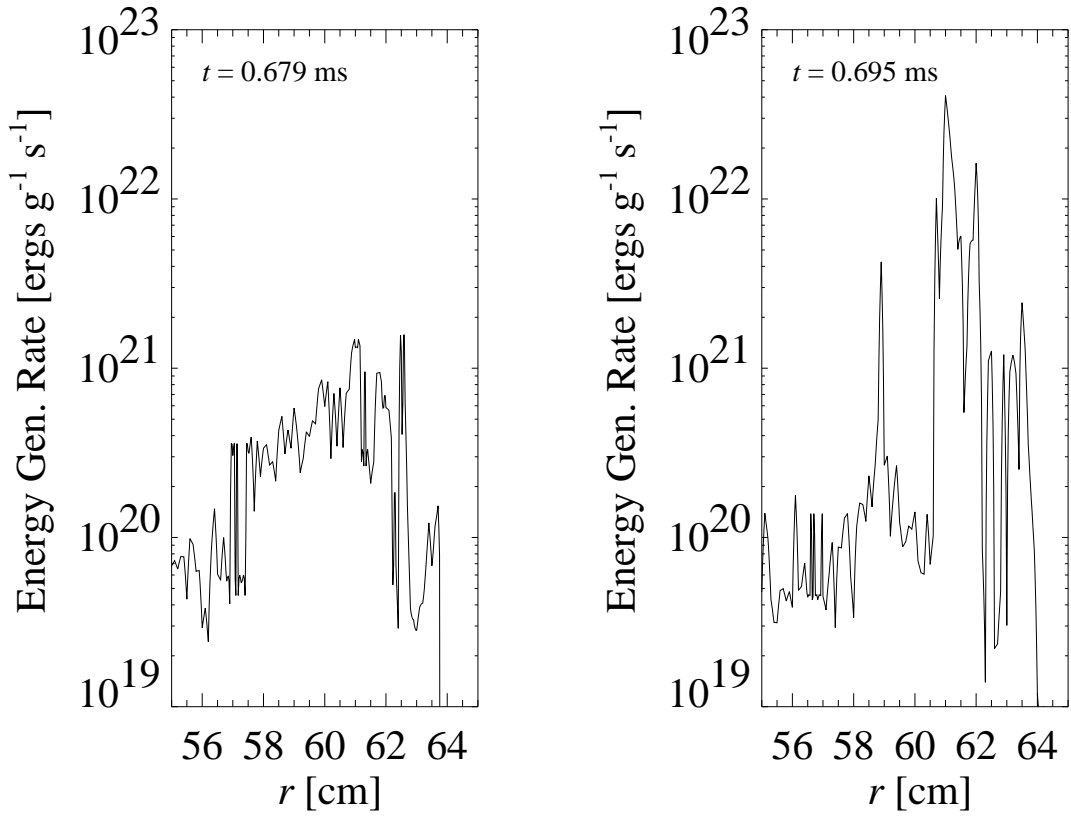}
\caption{A local explosion at the front of a turbulent flame
in the distributed regime. The profiles are taken from the model
A3}
\end{figure*}
%--------------------------
%eps
%--------------------------
\begin{figure*}[ht]
\figurenum{6}
\plotone{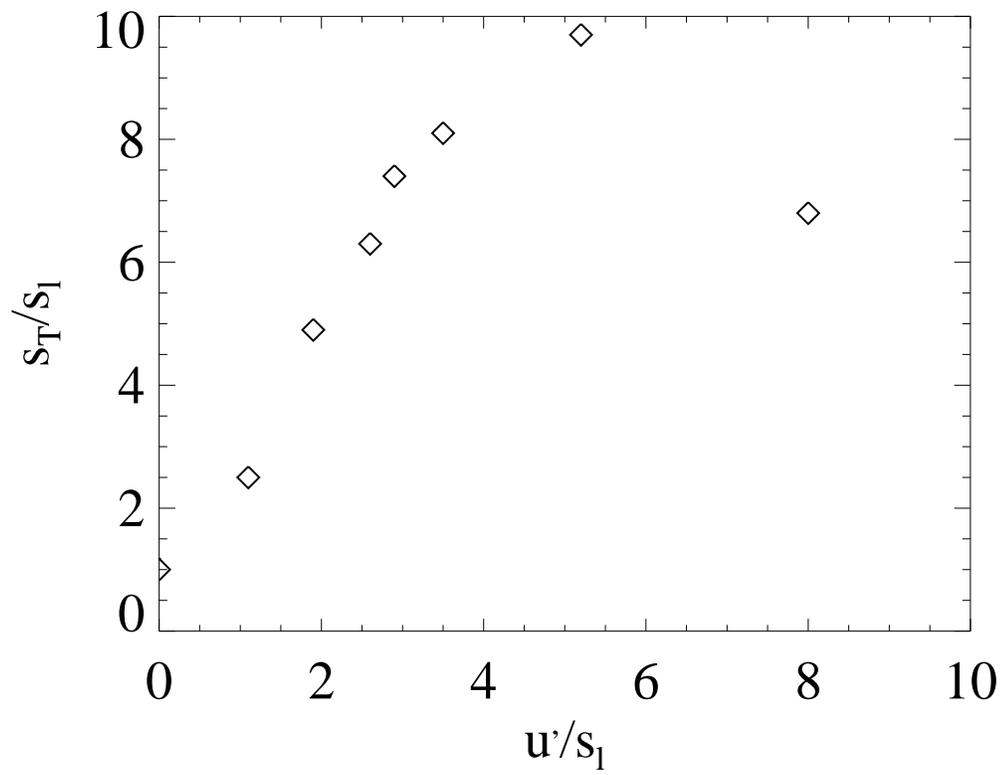}
\caption{Experimental data for the turbulent flame
speeds  taken from Abdel--Gayed et al. 1987.}
\end{figure*}

\end{document}